\documentclass [a4paper,twoside,12pt]{article}
\usepackage{a4wide,amsmath,amssymb,latexsym}
\usepackage{delarray,graphics,graphpap,alltt,graphicx}
\usepackage[latin1]{inputenc}       
\usepackage{algorithm}       
\usepackage{algorithmic}     
\usepackage {a4wide,amsmath,amssymb,latexsym,graphicx}
\usepackage{rotating}

\newcommand{\bfs}{\bfseries}   
\newcommand{\its}{\itshape}    

\newcommand{\dimbpd}{\dim_{\ensuremath{\mathsf{BPD}}}}

\newcommand{\dimfs}{\dim_{\ensuremath{\mathsf{FS}}}}

\newcommand{\dimH}{\dim_{\ensuremath{\mathsf{H}}}}

\newcommand{\robpd}{\rho _{\ensuremath{\mathsf{BPD}}}}

\newcommand{\roC}{\rho _C}

\newtheorem{theorem}{Theorem}[section]
\newtheorem{proposition}[theorem]{Proposition}

\newtheorem{lemma}[theorem]{Lemma}
\newtheorem{corollary}[theorem]{Corollary}

\newtheorem{remark}[theorem]{Remark}
\newtheorem{observation}[theorem]{Observation}

\newenvironment{definition}
{ {\noindent {\bf Definition.}} } {  }

\newenvironment{proposition*}[1]
{ {\noindent {\bf Proposition #1.}} } {  }

\newcommand{\bool}[1]{\{ 0,1\}^{#1}}

\newcommand{\N}{\mathbb{N}}

\def\squareforqed{\hbox{\rlap{$\sqcap$}$\sqcup$}}
\def\qed{\ifmmode\squareforqed\else{\unskip\nobreak\hfil
\penalty50\hskip1em\null\nobreak\hfil\squareforqed
\parfillskip=0pt\finalhyphendemerits=0\endgraf}\fi}

\newtheorem{construction}{CONSTRUCTION}[section]

\title{Bounded Pushdown dimension vs Lempel Ziv information density}

\author{Pilar Albert, Elvira Mayordomo, and Philippe Moser \footnote{Dept. de Informática e Ingeniería de Sistemas
, Universidad de Zaragoza. Edificio Ada Byron, María de Luna 1 -
E-50018 Zaragoza (Spain). Email: \{mpalbert, elvira\}@unizar.es and
mosersan@gmail.com. Research supported in part by Spanish Government
MEC
   Project TIN 2005-08832-C03-02, by Aragón Government Dept. Ciencia, Tecnología y Universidad,
   subvención destinada a la formación de personal
   investigador-B068/2006
 and by Spanish Government MEC Program Juan de la
   Cierva.}}
\date{}

\begin{document}
\maketitle

\begin{abstract}
    In this paper we introduce a variant of pushdown dimension called bounded pushdown (BPD) dimension,
    that measures the density of information contained in a sequence, relative to a BPD automata, i.e.
    a finite state machine equipped with an extra infinite memory stack, with the additional requirement that
    every input symbol only allows a bounded number of stack
    movements. BPD automata are a natural real-time restriction of
    pushdown automata.
    We show that BPD dimension is a
    robust notion by giving an equivalent characterization of BPD dimension in terms of BPD compressors.
    We then study the relationships between BPD compression, and the
    standard Lempel-Ziv (LZ) compression algorithm, and show that in contrast to the finite-state compressor
    case, LZ is not universal for bounded pushdown compressors in a strong sense: we construct a sequence
    that LZ fails to compress significantly, but that is compressed by at least a factor 2 by a
    BPD
    compressor. As a corollary we obtain a strong separation between finite-state and BPD dimension.

\end{abstract}

\section*{Keywords}

Information lossless compressors, finite state (bounded pushdown)
dimension, Lempel-Ziv compression algorithm.


\section{Introduction}

Effective versions of fractal dimension have been developed since
2000 \cite{DCC,DISS} and used for the quantitative study of
complexity classes, information theory and data compression, and
back in fractal geometry (see recent surveys in
\cite{efd,fgcc,efdait}). Here we are interested in information
theory and data compression, where it is known that for several
different bounds on the computing power,
 effective
dimensions capture what can be considered the inherent information
content of a sequence in the corresponding setting \cite{efdait}. In
the today realistic context of massive data streams we need to
consider very low resource-bounds, such as finite memory or
finite-time per input symbol.

 The finite state dimension of an infinite sequence \cite{Dai:2004:FSD}, is a measure of the  amount of randomness
    contained in the sequence within  a finite-memory setting. It is a robust quantity, that has been shown
    to admit several characterizations in terms of finite-state information lossless compressors
    (introduced by Huffman \cite{DBLP:Huffman_compressor}, \cite{Dai:2004:FSD}), finite-state decompressors
    \cite{DotMos06,ShLeZi91},
    finite-state predictors in the logloss model
    \cite{esdaicc}, and block entropy rates
    \cite{DBLP:journals/tcs/BourkeHV05}. It is an effectivization of the general notion of Hausdorff dimension at the level
    of finite-state machines.
    Informally, the finite state dimension assigns  every sequence a number $s\in[0,1]$, that characterizes the randomness density
    in the sequence (or equivalently its compression ratio), where the larger the dimension the more randomness is contained in the sequence.

    In a recent line of research, Doty and Nichols \cite{doty-2005} investigated a variant of finite-state dimension, where the
    finite state machine comes equipped with an  infinite  memory stack and is called a pushdown automata,
    yielding the notion of pushdown dimension.
    Hence the pushdown dimension of a sequence, is a measure of the density of randomness in the sequence as viewed by a
    pushdown automata. Since a finite-state automata is a special case of a pushdown automata, the pushdown dimension
    of a sequence is a lower bound for its finite state dimension. It was shown in \cite{doty-2005}, that there are sequences
    for which the pushdown dimension is at most half its finite state dimension, hence yielding a strong separation
    between the two notions. Unfortunately the notion of pushdown dimension is not known to enjoy any of
    the equivalent characterizations that finite state dimension
    does. Moreover, the computation time per input symbol can be
    unbounded, which rules out this model for many real-time
    applications.

    In this paper we introduce a variant of pushdown dimension called bounded pushdown (BPD) dimension:
    Whereas pushdown automata  can choose not to read their input and only
    work with their stack for as many steps as they wish (each such step is called a lambda transition),
    we add the additional real-time constraint that
    the sequences of lambda transitions are bounded, i.e.  we only allow a bounded number of stack movements  per each input
    symbol.

    We define the notion of bounded pushdown dimension as the
    natural effectivitation of Hausdorff dimension via Lutz's gale
    characterization \cite{DCC}. We provide evidence
    that bounded pushdown dimension is a robust notion by giving
    a compression characterization; i.e. we introduce BPD information-lossless compressors and show that
    the best compression ratio achievable on a sequence by BPD compressors is exactly its BPD dimension.

    In the context of compression, we  study the relationship between BPD compression and the
    standard Lempel-Ziv (LZ) compression algorithm \cite{DBLP:journals/tit/ZivL78}. It is well known that the LZ compression ratio
    of any sequence is a lower bound for its finite state compressibility \cite{DBLP:journals/tit/ZivL78}, i.e. LZ compresses every
    sequence at least as well as any finite-state information lossless compressor.
    We show that this fails dramatically in the context of BPD compressors, by constructing a sequence
    that LZ fails to compress significantly, but is compressed by at least a factor 2 by a
    BPD
    compressor, thus yielding a strong separation between LZ and BPD dimension.   This
    implies that we have
    the same separation between LZ and (unbounded) pushdown dimension, and between finite state dimension
    \cite{Dai:2004:FSD} and  BPD dimension.

Section 2 contains the preliminaries, section 3 presents BPD
dimension and its basic properties, section 4 proves the equivalence
of BPD compression and dimension and section 5 contains the
separation of BPD compression from Lempel Ziv compression. The
proofs are postponed to the appendix.


\section{Preliminaries}

We write $\mathbb{Z}$ for the set of all integers, $\mathbb{N}$ for
the set of all nonnegative integers and $\mathbb{Z^+}$ for the set
of all positive integers. Let $\Sigma$ be a finite alphabet, with
$|\Sigma| \geq 2$. $\Sigma^*$ denotes the set of finite strings, and
$\Sigma ^\infty $ the set of infinite sequences. We write $|w|$ for
the length of a string $w$ in $\Sigma ^*$. The empty string is
denoted $\lambda $. For $S$ $\in $ $\Sigma ^\infty $ and $i,j$ $\in
$ $\mathbb{N}$, we write $S[i..j]$ for the string consisting of the
$i^{\textrm{th}}$ through $j^{\textrm{th}}$ symbols of $S$, with the
convention that $S[i..j]=\lambda $ if $i>j$, and $S[0]$ is the
leftmost symbol of $S$. We write $S[i]$ for $S[i..i]$ (the
$i^{\textrm{th}}$ symbol of $S$). For $w$ $\in $ $\Sigma ^*$ and $S$
$\in $ $\Sigma ^\infty $, we write $w\sqsubseteq S$ if $w$ is a
prefix of $S$, i.e., if $w=S[0..|w|-1]$. All logarithms are taken in
base $|\Sigma|$.


\section{Bounded Pushdown Dimension}


In this section we first recall Lutz's characterization of Hasudorff
dimension in terms of gales that can be used to effectivize
dimension. Then we  introduce Bounded Pushdown dimension based on
the concept of BPD gamblers and give its basic properties.

\begin{definition}\cite{DCC} Let $s \in [0,\infty ).$
\begin{enumerate}
  \item An {\its s-gale} is a function $d:\Sigma ^*\rightarrow [0,\infty
  )$ that satisfies the condition
    \begin{equation}\label{equ:gale}d(w)=\frac{\sum\limits_{a \in \Sigma } d(wa)}{|\Sigma |^{s}
    }\end{equation}
for all $w \in \Sigma^*.$
  \item A {\its martingale} is a 1-gale.
\end{enumerate}
\end{definition}

Intuitively, an $s$-gale is a strategy for betting on the successive
symbols of a sequence $S \in \Sigma^\infty$. For each prefix $w$ of
$S$, $d(w)$ is the capital (amount of money) that $d$ has after
having bet on $S[0..|w|-1]$. When betting on the next symbol $b$ of
a prefix $wb$ of $S$, assuming symbol $b$ is equally likely to be
any value in $\Sigma $, equation (\ref{equ:gale}) guarantees that
the expected value of $d(wb)$ is $|\Sigma|^{-1}\sum\limits_{a \in
\Sigma} d(wa)=|\Sigma |^{s-1}d(w)$. If $s=1$, this expected value is
exactly $d(w)$, so the payoffs are ``fair''.

\begin{definition} Let $d$ be an $s$-gale, where $s \in [0,\infty)$.
\begin{enumerate}
  \item We say that $d$ {\itshape succeeds} on a sequence $S
  \in \Sigma ^\infty$ if
\begin{center}
$\limsup\limits_{n\rightarrow \infty }d(S[0..n-1])=\infty .$
\end{center}

\item The {\itshape success set} of $d$ is
   \begin{center}
    $S^\infty [d]=\{S \in \Sigma ^\infty \mid d$ succeeds on $S\}.$
   \end{center}

\end{enumerate}
\end{definition}

\begin{observation}\label{gales-obs} Let $s,s'$ $\in$ $[0,\infty )$. For
every $s$-gale $d$, the function $d':\Sigma^*\rightarrow [0,\infty
)$ defined by $d'(w)=|\Sigma |^{(s'-s)|w|}d(w)$ is an $s'$-gale.
Moreover, if $s\leq s'$, then $S^\infty [d]\subseteq S^\infty [d']$.
\end{observation}

Lutz characterized Hausdorff dimension using gales as follows.

\begin{theorem}\cite{DCC}
Given a set  $X\subseteq\Sigma^\infty$, if $\dimH(X)$ is the
Hausdorff dimension of $X$ \cite{Falc85}, then
\[\dimH(X)=\inf\{s\,|\, \mbox{there is an }s-\mbox{gale }d \mbox{
such that }X\subseteq S^\infty [d]\}\]\end{theorem}

The idea for a Bounded Pushdown dimension is to consider only
$s$-gales that are computable by a Bounded Pushdown (BPD) gambler.
Bounded Pushdown gamblers are finite-state gamblers
\cite{Dai:2004:FSD} with an extra memory stack, that is used both by
the transition and betting functions. Additionally, BPDG's are
allowed to delay reading the next character of the input --they read
$\lambda $ from the input-- in order to alter the content of their
stack, but they cannot do this more than a constant number of times
per each input symbol. During such $\lambda $-transitions, the
gambler's capital remains unchanged.

The betting function returns a probability measure over the input
alphabet.

\begin{definition}
Let $\Sigma $ be a finite alphabet. $\Delta _\mathbb{Q}(\Sigma )$ is
the set of all rational-valued probability measures over $\Sigma $,
i.e., all functions $\pi :\Sigma \longrightarrow [0,1]\cap
\mathbb{Q}$ such that $\sum\limits_ {a \in \Sigma}\pi (a)=1$.
\end{definition}


We are ready to define BPD gamblers.

\begin{definition}
A {\sl bounded pushdown gambler (BPDG)\/} is an 8-tuple $G=$($Q$,
$\Sigma $, $\Gamma $, $\delta $, $\beta $, $q_0$, $z_0$, $c$) where
\begin{itemize}
\item $Q$ is a finite set of {\itshape states},
\item $\Sigma $ is the finite input alphabet,
                   \item $\Gamma$ is the finite {\itshape stack
                   alphabet},
                   \item $\delta :Q\times (\Sigma \cup \{\lambda \})\times \Gamma\rightarrow Q\times \Gamma
                   ^*$ is the {\itshape transition function} (for simplicity we use
                   the notation $\delta (q,b,a)=\bot $ when
                   undefined; and we write $\delta (q,b,a)=(\delta _Q(q,b,a),\delta _{\Gamma
                   ^*}(q,b,a))$),
                   \item $\beta :Q\times \Gamma \rightarrow
                   \Delta _\mathbb{Q}(\Sigma)$ is the {\itshape betting function},
                   \item $q_0$ $\in $ $Q$ is the {\itshape start
                   state},
                   \item $z_0$ $\in $ $\Gamma $ is the {\itshape start stack
                   symbol},
\item $c$ $\in$ $\mathbb{N}$ is a constant such that  the number of $\lambda
$-transitions per input symbol is at most $c$,
                   \end{itemize}
                   with the two additional restrictions:
                   \begin{enumerate}
\item                   for each $q$ $\in $ $Q$ and $a$ $\in $ $\Gamma $ at least one of the following
holds\begin{itemize}
            \item $\delta (q,\lambda ,a)=\perp $
            \item $\delta (q,b,a)=\perp $ for all $b$ $\in $ $\Sigma$
            \end{itemize}
\item for every $q$ $\in $ $Q$, $b$ $\in $ $\Sigma \cup \{\lambda \}$,
either $\delta (q,b,z_0)=\perp $, or $\delta (q,b,z_0)=(q',vz_0)$,
where $q'$ $\in $ $Q$ and $v$ $\in $ $\Gamma ^*$.
\end{enumerate}
We denote with $BPDG$  the set of all bounded pushdown gamblers.
\end{definition}

The transition function $\delta $ outputs a new state and a string
$z'$ $\in $ $\Gamma ^*$. Informally, $\delta (q, w, a)=(q', z')$
means that in state $q$, reading input $w$, and popping symbol $a$
from the stack, $\delta$ enters state $q'$ and pushes $z'$ to the
stack.

Note that $w$ can be $\lambda$ (ie, a $\lambda$-transition: the
input is ignored and $\delta$ only computes with the stack) but this
only happens at most $c$ times per input symbol. Any pair (state,
stack symbol) can either be a
                                  $\lambda$-transition pair or a
                                  non $\lambda$-transition pair
                                  exclusively, because the first additional restriction
                                  enforces
                                  determinism.

                                Moreover, since $z_0$ represents the bottom of the stack, we
restrict $\delta $ so that $z_0$ cannot be removed from the bottom
by the second additional restriction.

We can extend $\delta $ in the usual way to
$$\delta ^*:Q\times (\Sigma \cup \{\lambda \})\times \Gamma ^+
\rightarrow Q\times \Gamma ^*,$$ where for all $q$ $\in $ $Q$, $a$
$\in $ $\Gamma $, $v$ $\in $ $\Gamma ^*$, and $b$ $\in $ $\Sigma
\cup \{\lambda \}$
$$\delta ^*(q,b,av)=\left\{
                         \begin{array}{ll}
                           (\delta _Q(q,b,a),\delta _{\Gamma ^*}(q,b,a)v) & \hbox{if $\delta (q,b,a)\neq \perp $,} \\
                           \perp  & \hbox{otherwise.}
                         \end{array}
                       \right.
    $$
We denote $\delta ^*$ by $\delta $.

 For each $i\geq 2$, we will use the notation
$$\delta ^{i}(q, \lambda , v)=\delta (\delta ^{i-1}_Q(q,\lambda ,v),
\lambda , \delta ^
    {i-1}_{\Gamma ^*}(q,\lambda ,v))$$
where $$\delta ^{1}(q, \lambda , v)=\delta (q, \lambda , v).$$ Since
$\delta$ is $c$-bounded we have that for any  $q$ $\in $ $Q$, $v$
$\in $ $\Gamma ^*$,
    $$\delta ^{c+1}(q,\lambda ,v)=\bot $$

We also consider the extended transition function
$$\delta ^{**}:Q\times \Sigma^* \times \Gamma
^+\rightarrow Q\times \Gamma
    ^*,$$
defined for all $q$ $\in $ $Q$, $a$ $\in $ $\Gamma $, $v$ $\in $
$\Gamma ^*$, $w$ $\in $ $\Sigma^*$, and $b$ $\in $ $\Sigma$ by
$$\delta ^{**}(q, \lambda , av)=(q,av)$$
$$\delta ^{**}(q, wb, av)=\delta (\delta ^{i}_Q(\widetilde{q},
\lambda , \widetilde{a}
    \widetilde{v}), b, \delta ^{i}_{\Gamma ^*}(\widetilde{q}, \lambda , \widetilde{a} \widetilde{v}))$$
if $\delta ^{**}(q, w, av)=(\widetilde{q}, \widetilde{a}
\widetilde{v})$, $\delta ^{i}(\widetilde{q}, \lambda ,
\widetilde{a}\widetilde{v}) \neq \perp $ and $\delta
^{i+1}(\widetilde{q}, \lambda , \widetilde{a}\widetilde{v})= \perp
$, $i\leq c$.

 That is, $\lambda $-transitions are inside the definition of $\delta ^{**}(q, b, av)$,
for $b$ $\in $ $\Sigma$. Notice that $\delta ^{**}$ is not defined
on an empty stack string, therefore $av$ needs to be long enough in
order that $\delta ^{**}(q, b, av)\neq \perp $.

 We denote $\delta ^{**}$ by $\delta $, and $\delta
(q_0,w,z_0)$ by $\delta (w)$. We write $\delta =(\delta _Q,\delta
_{\Gamma^*})$ for simplicity.

 We also consider the usual extension of $\beta $
$$\beta ^*:Q\times \Gamma ^+\rightarrow \Delta_\mathbb{Q}(\Sigma ),$$
defined for all $q$ $\in $ $Q$, $a$ $\in $ $\Gamma$, and $v$ $\in $
$\Gamma ^*$ by
$$\beta ^*(q,av)=\beta (q,a),$$
and denote $\beta ^*$ by $\beta $.

We use  BPDG to compute martingales. Intuitively, suppose a BPDG $G$
is to bet on sequence $S$ has already bet on $w\sqsubset S$, with
current capital $x$ $\in$ $\mathbb{Q}$, current state $q$ $\in$ $Q$
and current top stack symbol $a$. Then for $b$ $\in$ $\Sigma$, $G$
bets the quantity $x\beta (q, a)(b)$ of its capital that the next
symbol of $S$ is $b$. If the bet is correct (that is, if
$wb\sqsubset S$) and since payoffs are fair, $G$ has capital
$|\Sigma |x\beta (q, a)(b)$. Formally,

\begin{definition} Let $G=(Q, \Sigma , \Gamma , \delta , \beta , q_0,
z_0, c)$ be a bounded pushdown gambler. The {\itshape martingale} of
$G$ is the function
$$d_G:\Sigma ^*\rightarrow [0,\infty )$$
defined by the recursion
$$d_G(\lambda )=1$$
$$d_G(wb)=|\Sigma |d_G(w)\beta (\delta (w))(b)$$
for all $w$ $\in $ $\Sigma ^*$ and $b$ $\in $ $\Sigma $.
\end{definition}

By Observation \ref{gales-obs}, a BPDG $G$ actually yields an
$s$-gale for every $s \in [0,\infty )$. We call it the $s$-gale of
$G$, and denote it by $$d^s_G(w)=|\Sigma |^{(s-1)|w|}d_G(w).$$ A
bounded pushdown $s$-gale is an $s$-gale $d$ for which there exists
a BPDG such that $d^s_G=d$.

The first two properties of BPD gamblers are that any number of
$\lambda$-transitions can be replaced by a single
$\lambda$-transition and that the stack alphabet does not give
additional power.

\begin{proposition}\label{c-1gamblers}
Let $G=(Q, \Sigma , \Gamma , \delta , \beta , q_0, z_0, c)$ be a
BPDG. Then there is a BPDG $G'=(Q', \Sigma , \Gamma' , \delta' ,
\beta' , q'_0, z'_0, 1)$ such that $d_G=d_{G'}$.\end{proposition}

From now on we shall
  assume that the maximum number of $\lambda$-transitions $c$ is 1.

\begin{proposition}\label{stackal}
Let $G=(Q, \Sigma , \Gamma , \delta , \beta , q_0, z_0, c)$ be a
BPDG. Then there is a BPDG $G'=(Q', \Sigma , \{0, 1, z'_0\} ,
\delta' , \beta' , q'_0, z'_0, c')$ such that
$d_G=d_{G'}$.\end{proposition}

 Let us define bounded pushdown
dimension. Intuitively, the BPD dimension of a sequence is the
smallest $s$ such that there is a BPD-$s$-gale that succeeds on the
sequence.

\begin{definition} The {\itshape bounded pushdown dimension} of a
set $X\subseteq \Sigma ^\infty$ is
$$\dimbpd(X)= \inf\{s\,|\, \mbox{there is a bounded pushdown }s-\mbox{gale }d \mbox{
such that }X\subseteq S^\infty [d]\}.$$
\end{definition}


\section{Dimension and compression}

In this section we characterize the bounded pushdown dimension of
individual sequences in terms of bounded pushdown compressibility,
therefore BPD dimension is a natural and robust definition.

\begin{definition}A {\itshape bounded pushdown compressor} $(BPDC)$
is an 8-tuple
$$C=(Q, \Sigma , \Gamma , \delta , \nu , q_0, z_0, c)$$
where\begin{itemize}
\item $Q$ is a finite set of states,
\item $\Sigma$ is the finite input and output alphabet,
       \item $\Gamma $ is the finite stack alphabet,
       \item $\delta :Q\times (\Sigma \cup \{\lambda \})\times \Gamma \rightarrow Q\times \Gamma
       ^*$ is the transition function,
       \item $\nu :Q\times \Sigma \times \Gamma \rightarrow
       \Sigma ^*$ is the output function,
       \item $q_0$ $\in $ $Q$ is the initial state,
       \item $z_0$ $\in $ $\Gamma $ is the start stack symbol,
       \item $c$ $\in$ $\mathbb{N}$ is a constant such that  the number of $\lambda
$-transitions  per input symbol is at most $c$,
     \end{itemize}
      with the two additional restrictions:
                   \begin{enumerate}
\item                   for each $q$ $\in $ $Q$ and $a$ $\in $ $\Gamma $ at least one of the following
holds\begin{itemize}
            \item $\delta (q,\lambda ,a)=\perp $
            \item $\delta (q,b,a)=\perp $ for all $b$ $\in $ $\Sigma$
            \end{itemize}
\item for every $q$ $\in $ $Q$, $b$ $\in $ $\Sigma \cup \{\lambda \}$,
either $\delta (q,b,z_0)=\perp $, or $\delta (q,b,z_0)=(q',vz_0)$,
where $q'$ $\in $ $Q$ and $v$ $\in $ $\Gamma ^*$.
\end{enumerate}
\end{definition}

We extend $\delta $ to $\delta ^{**}:Q\times \Sigma ^*\times \Gamma
^+\rightarrow Q\times \Gamma ^*$ as before, and denote $\delta
^{**}$ by $\delta $ and $\delta (q_0,w,z_0)$ by $\delta (w)$.

 For $q$ $\in $ $Q$, $w$ $\in $ $\Sigma ^*$ and $z$
$\in $ $\Gamma ^+$, we define the {\itshape output} from state $q$
on input $w$ reading $z$ on the top of the stack to be the string
$\nu ^*(q,w,z)$ (denoted by $\nu (q,w,z)$) with
$$\nu (q,\lambda ,z)= \lambda $$
$$\nu (q,wb,z)=\nu (q,w,z)\nu (\delta _Q(q,w,z), b, \delta _{\Gamma ^*}(q,w,z))$$
for $w$ $\in $ $\Sigma ^*$ and $b$ $\in $ $\Sigma $. We then define
the {\itshape output} of $C$ on input $w$ $\in $ $\Sigma ^*$ to be
the string
$$C(w)=\nu (q_0,w,z_0).$$

We can restrict $\lambda$-transitions to a single one and the stack
alphabet to three symbols.

\begin{proposition}\label{c-1compressors}
Let $C=(Q, \Sigma, \Gamma, \delta, \nu, q_0, z_0, c)$ be a BPDC.
Then there is a BPDC $C'=(Q', \Sigma , \Gamma' , \delta' , \nu' ,
q'_0, z'_0, 1)$ such that $C(w)=C'(w)$ for every
$w\in\Sigma^*$.\end{proposition}

\begin{proposition}
Let $C=(Q, \Sigma, \Gamma, \delta, \nu, q_0, z_0, c)$ be a BPDC.
Then there is a BPDC $C'=(Q', \Sigma , \{0, 1, z'_0\} , \delta' ,
\nu' , q'_0, z'_0, c')$ such that $C(w)=C'(w)$ for every
$w\in\Sigma^*$.\end{proposition}

We are interested in   \textit{information lossless} compressors,
that is, $w$ must be recoverable from $C(w)$ and the final state.

\begin{definition} A BPDC $C=(Q, \Sigma , \Gamma
, \delta , \nu , q_0, z_0)$ is {\itshape information-lossless}
($IL$) if the function
$$\Sigma ^*\rightarrow \Sigma ^*\times Q$$
$$w\rightarrow (C(w),\delta _Q(w))$$
is one-to-one. An {\itshape information-lossless bounded pushdown
compressor} ($ILBPDC$) is a BPDC that is IL.
\end{definition}

 Intuitively, a BPDC {\itshape compresses} a string $w$ if $|C(w)|$
is significantly less than $|w|$. Of course, if $C$ is $IL$, then
not all strings can be compressed. Our interest here is in the
degree (if any) to which the prefixes of a given sequence $S$ $\in $
$\Sigma ^\infty $ can be compressed by an ILBPDC.

\begin{definition} If $C$ is a BPDC and $S$ $\in $ $\Sigma ^\infty $, then
the {\itshape compression ratio} of $C$ on $S$ is
$$\roC(S)=\liminf\limits_{n\rightarrow \infty
    }\frac{|C(S[0..n-1])|}{n}.$$
\end{definition}

The BPD compression ratio of a sequence is the best compression
ratio achievable by an ILBPDC, that is

\begin{definition} The {\itshape bounded pushdown compression ratio}
of a sequence $S$ $\in $ $\Sigma ^\infty $ is
$$\robpd(S)=\inf \{\roC(S)\mid \text{ C is a ILBPDC}\}.$$
\end{definition}

The main result in this section states that the BPD dimension of a
sequence and its ILBPD compression ratio are the same, therefore BPD
dimension is the natural concept of density of information in the
BPD setting.

\begin{theorem} \label{main-theorem} For all $S$ $\in $ $\Sigma ^\infty $,
$$\dimbpd (S)=\robpd(S).$$
\end{theorem}


\section{Separating LZ from BPD}

In this section we prove that BPD compression can be much better
than the compression attained with the celebrated Lempel-Ziv
algorithm.

We start with a brief description of the LZ algorithm
\cite{DBLP:journals/tit/ZivL78}.

 We finish relating BPD dimension (and
compression) with the Lempel-Ziv algorithm.  Given an input
$x\in\Sigma^*$, LZ parses $x$ in different phrases $x_i$,
    i.e., $x=x_1 x_2 \ldots x_n$ ($x_i\in\Sigma^*$) such that
    every prefix $y\sqsubset x_i$, appears before $x_i$ in the parsing (i.e. there exists $j<i$ s.t. $x_j=y$).
    Therefore for every $i$, $x_i = x_{l(i)}b_i$ for $l(i)<i$ and
    $b_i\in\Sigma$. We sometimes denote the number of phrases in the
    parsing of $x$ as $C(x)$.

    LZ encodes $x_i$ by a prefix free encoding of $l(i)$ and the
    symbol $b_i$, that is, if $x=x_1 x_2 \ldots x_n$ as before, the
    output of LZ on input $x$ is
    \[LZ(x)= c_{l(1)}b_1 c_{l(2)}b_2 \ldots c_{l(n)}b_n\] where
    $c_i$ is a prefix-free coding of $i$ (and $x_0=\lambda$).

    LZ is usually restricted to the binary alphabet, but the
    description above is valid for any $\Sigma$.

    For a sequence $S\in\Sigma^{\infty}$, the LZ compression ratio is given by
    $$
        \rho_{LZ}(S) = \liminf_{n\rightarrow\infty} \frac{|LZ(S[0\ldots n-1])|}{n}        .
    $$
It is well known that LZ \cite{DBLP:journals/tit/ZivL78} yields a
lower bound on the finite-state dimension (or finite-state
compressibility) of a sequence \cite{DBLP:journals/tit/ZivL78}, ie,
LZ is universal for finite-state compressors.

The following result shows that this is not true for BPD (hence PD)
dimension, in a strong sense: we construct a sequence $S$ that
cannot be compressed by LZ, but that has BPD compression ratio less
than $\frac{1}{2}$.

\begin{theorem} \label{LZ}
For every $m\in\N$, there is a sequence $S \in \{0,1\}^\infty$ such
that $$\rho_{LZ}(S)
> 1 - \frac{1}{m}$$ and $$\dimbpd(S)\leq \frac{1}{2}.$$
\end{theorem}

As a corollary we obtain a separation of finite-state dimension and
bounded pushdown dimension. A similar result between finite-state
dimension and pushdown dimension was proved in \cite{doty-2005}.

\begin{corollary}
For any $m \in \mathbb{N}$, there exists a sequence $S \in
\{0,1\}^\infty$ such that $$\dimfs(S)>1-\frac{1}{m}$$ and
$$\dimbpd(S)\leq \frac{1}{2}.$$
\end{corollary}

\section*{Conclusion}
We have introduced Bounded Pushdown dimension, characterized it with
compression and compared it with Lempel-Ziv compression.  It is open
if there is a BPD compressor that is universal for Finite-State
compressors, which is true for the Lempel-Ziv algorithm,  and
whether Lempel-Ziv compression can surpass BPD-compression for some
sequence.


\newpage


\renewcommand{\thesection}{\Alph{section}}
\setcounter{section}{0} \setcounter{page}{1}


\pagestyle{plain}

 \vspace{8in}
\begin{center}
{\bf{\huge Technical Appendix}}
\end{center}

This appendix is devoted to proving Theorem \ref{main-theorem} and
Theorem \ref{LZ}. For the first one, we need the following:
\section{Proof of Theorem \ref{main-theorem}}

\begin{definition} A BPDG $G=(Q, \Sigma , \Gamma , \delta , \beta , q_0,
z_0)$ is {\itshape nonvanishing} if $0<\beta (q,z)(b)<1$ for all $q$
$\in $ $Q$, $b$ $\in $ $\Sigma $ and $z$ $\in $ $\Gamma $.
\end{definition}

\begin{lemma} \label{gamblers-lemma}For every BPDG $G$ and each $\varepsilon
>0$, there is a nonvanishing BPDG $G'$ such that for all $w$ $\in $
$\Sigma ^*$, $d_{G'}(w)\geq |\Sigma |^{-\varepsilon |w|}d_G(w)$.
\end{lemma}

{\bfs Proof of Lemma \ref{gamblers-lemma} .} Let $G=(Q, \Sigma ,
\delta , \beta , q_0, \Gamma , z_0)$ be a BPDG, and let $\varepsilon
> 0$. For each $q$ $\in $ $Q$, $z$ $\in $ $\Gamma $, $b$ $\in $
$\Sigma $,
$$1 - |\Sigma |^{-\varepsilon }\sum\limits_{b \in \Sigma}\beta (q,z)(b)= 1 - |\Sigma |^{-\varepsilon } >
    0,$$
so we can fix a rational $\beta '(q,z)(b)$ such that
$$|\Sigma |^{-\varepsilon }\beta (q,z)(b) < \beta '(q,z)(b) < 1 - |\Sigma |^{-\varepsilon }
    \sum\limits_{a \in \Sigma, a\neq b}\beta (q, z)(a)$$
and
$$\sum\limits_{b \in \Sigma} \beta '(q,z)(b)=1.$$
Then, $0 < \beta '(q,z)(b) < 1$ for each $q$ $\in $ $Q$, $b$ $\in $
$\Sigma$ and $z$ $\in $ $\Gamma $, therefore the BPDG $G'=(Q, \Sigma
, \delta , \beta ', q_0, \Gamma , z_0)$ is nonvanishing.

 Also, for all $q$ $\in $ $Q$, $b$ $\in $ $\Sigma$, $z$ $\in $ $\Gamma $,
$$\beta '(q,z)(b)\geq |\Sigma |^{-\varepsilon }\beta (q,z)(b)$$
so for all $w$ $\in $ $\Sigma ^*$, $d_{G'}(w)\geq |\Sigma
|^{-\varepsilon |w |}d_G(w)$.
\begin{flushright}
    $\Box $
\end{flushright}

{\bfseries Proof of Theorem \ref{main-theorem}} Let $S$ $\in $
$\Sigma ^\infty $. For each $n$ $\in $ $\mathbb{N}$, let
$w_n=S[0..n-1]$.

 To see that $\dimbpd(S)\leq \robpd(S)$, let $s>s'>\robpd(S)$. It suffices to show that $\dimbpd(S)\leq s$. By our choice of
$s'$, there is an 1-ILBPDC $C=(Q, \Sigma , \Gamma ,\delta , \nu ,
q_0, z_0)$ for which the set
$$I=\{n\in \mathbb
    {N}\mid |C(w_n)|<s'n\}$$
is infinite.

\begin{construction} \label{constructioncg} Given a $1$-bounded pushdown compressor (BPDC) \newline
$C=(Q, \Sigma , \Gamma
, \delta , \nu , q_0, z_0)$, and $k$ $\in$ $\mathbb{Z}^+$ , we
construct the $1$-bounded pushdown gambler (BPDG) $G=G(C,k)=(Q',
\Sigma , \Gamma ', \delta
', \beta ', q_0', z_0')$ as follows:\\
\\i) $Q'=Q\times \{0,1,\dots,k-1\}$\\
\\ii) $q_0'=(q_0,0)$\\
\\iii) $\Gamma '=\bigcup\limits_{i=2k}^{4k-1} \Gamma ^{i}$\\
\\iv) $z_0'=z_0^{2k}$\\
\\v) $\forall (q,i)\in Q', b\in \Sigma , a\in \Gamma ',$
$$\delta '((q, i), b, a)=\bigg (\Big (\delta _Q(q, b, \overline{a}),
(i+1)\bmod k\Big ), \widehat{\delta _{\Gamma ^*}(q, b,
\overline{a})}\bigg )$$ where for each $z$ $\in $ $(\Gamma ')^+$,
$\overline{z}$ $\in $ $\Gamma ^+$ is the $\Gamma $-string obtained
by concatenating the symbols of $z$, and for each $y$ $\in $ $\Gamma
^+$, if $y=y_1y_2\cdots y_{2kl+n}$ with $n<2k$, then $\widehat{y} $
$\in $ $(\Gamma ')^+ $ is such that $\widehat{y}
_1=y_1\cdots y_{2k+n}$, $\widehat{y} _2=y_{2k+n+1}\cdots y_{4k+n}$, $\dots$, $\widehat{y} _l=y_{2k(l-1)+n+1}\cdots y_{2kl+n}$.\\
\\vi) $\forall (q,i)\in Q', a\in \Gamma
', b \in \Sigma $
$$\beta '((q,i),a)(b)=\frac{\sigma
(q,b\Sigma ^{k-i-1},a)}{\sigma(q,\Sigma ^{k-i},a)}$$ where $\sigma
(q,A,a)=\sum\limits_{x\in A}|\Sigma |^{-|\nu (q,x,\overline{a})|}$ .
\end{construction}

\begin{lemma}\label{lemma1} In Construction \ref{constructioncg}, if $|w|$ is a multiple
of $k$ and $u$ $\in $ $\Sigma ^{\leq k}$, then
$$d_G(wu)=|\Sigma |^{|u| - |\nu (\delta _Q(w), u, \delta _{\Gamma ^*}(w))|}
    \frac{\sigma (\delta _Q(wu),\Sigma ^{k-|u|}, \widehat{\delta _{\Gamma ^*}(wu)})}
    {\sigma (\delta _Q(w),\Sigma ^k, \widehat{\delta _{\Gamma
    ^*}(w)})}d_G(w).$$
\end{lemma}

{\bfseries Proof of Lemma \ref{lemma1}.} We use induction on the
string $u$. If $u=\lambda $, the lemma is clear. Assume that it
holds for $u$, where $u$ $\in $ $\Sigma ^{<k}$, and let $b$ $\in $
$\Sigma $. Then
\begin{equation*}
\begin{aligned}
    d_G(wub)&=|\Sigma |\frac{\sigma (\delta _Q(wu),b\Sigma ^{k-|u|-1}, \widehat{\delta _{\Gamma ^*}(wu)})}
    {\sigma (\delta _Q(wu),\Sigma ^{k-|u|}, \widehat{\delta _{\Gamma ^*}(wu)})}d_G(wu)\\
    &=|\Sigma |^{1-|\nu (\delta _Q(wu),b, \delta _{\Gamma ^*}(wu))|}
    \frac{\sigma (\delta _Q(wub),\Sigma ^{k-|u|-1}, \widehat {\delta _{\Gamma ^*}(wub)})}
    {\sigma (\delta _Q(wu),\Sigma ^{k-|u|}, \widehat{\delta _{\Gamma ^*}(wu)})}d_G(wu)
\end{aligned}
\end{equation*}
so by the induction hypothesis the lemma holds for $ub$.
\begin{flushright}
$\Box $
\end{flushright}

\begin{lemma} \label{lemma2}In Construction \ref{constructioncg}, if $w=w_0w_1\cdots w_{n-1}$, where each
$w_i$ $\in $ $\Sigma ^k$ , then
$$d_G(w)=\frac{|\Sigma |^{|w| - |C(w)|}}{\prod\limits_{i=0}^{n-1} \sigma (\delta _Q(w_0\cdots w_{i-1}),\Sigma ^k, \widehat{\delta _{\Gamma
^*}(w_0\cdots w_{i-1})})}.$$
\end{lemma}

{\bfseries Proof of Lemma \ref{lemma2}.} We use induction on $n$.
For $n=0$, the identity is clear. Assume that it holds for
$w=w_0w_1\cdots w_{n-1}$, with each $w_i$ $\in $ $\Sigma ^k$, and
let $w'=w_0w_1\cdots w_n$. Then Lemma \ref{lemma1} with $u=w_n$
tells us that
$$d_G(w')=\frac{|\Sigma | ^{k-|\nu (\delta _Q(w), w_n, \delta _{\Gamma ^*}(w))|}}
    {\sigma (\delta _Q(w), \Sigma ^k, \widehat{\delta _{\Gamma
    ^*}(w)})}d_G(w)$$
whence the identity holds for $w'$ by the induction hypothesis.
\begin{flushright}
    $\Box $
\end{flushright}

\begin{lemma}\label{lemma3} In Construction \ref{constructioncg}, if $C$ is IL and $|w|$
is a multiple of $k$, then
$$d_G(w)\geq |\Sigma | ^{|w| - |C(w)| -\frac{|w|}{k} (l + \log m + \log k +
    1)},$$
where $l=\lceil \log |Q|\rceil $ and $m=$ $\max \{|\nu (q,b,a)|\mid
q \in Q, b \in \Sigma , a \in \Gamma ^2 \}$.
\end{lemma}

{\bfseries Proof of Lemma \ref{lemma3}.} We prove that for each $z$
$\in $ $\Sigma ^*$, \[\sigma (\delta _Q(z),\Sigma ^k,
\widehat{\delta _{\Gamma ^*}(z)})\leq |\Sigma | ^{l+\log m+\log
k+1}.\]

 To see this, fix $z$ $\in $ $\Sigma ^*$ and observe that at
most $|Q|$ strings $w$ $\in $ $\Sigma ^k$ can have the same output
from state $\delta _Q(z)$ with stack content $\delta _{\Gamma
^*}(z)$. Therefore, the number of $w$ $\in $ $\Sigma ^k$ for which
$|\nu (\delta _Q(z), w, \delta _{\Gamma ^*}(z))|=j$ does not exceed
$|Q||\Sigma | ^j$. Hence
\begin{equation*}
\begin{aligned}
    \sigma (\delta _Q(z),\Sigma ^k, \widehat{\delta _{\Gamma ^*}(z)})
    &=\sum\limits_{w\in \Sigma ^k} |\Sigma | ^{-|\nu (\delta _Q(z), w, \delta _{\Gamma ^*}(z))|}\leq \sum
    \limits_{j=0}^{mk} |Q||\Sigma | ^j|\Sigma | ^{-j}=|Q|(mk+1)\\
    &\leq |\Sigma | ^{l+\log m+\log k+1}.
\end{aligned}
\end{equation*}
It follows by Lemma \ref{lemma2} that
$$d_G(w)=|\Sigma | ^{|w|-|C(w)|-\frac{|w|}{k}(l+\log m+\log k+1)}.$$
\begin{flushright}
    $\Box $
\end{flushright}

\begin{lemma} \label{lemma4}In Construction \ref{constructioncg}, if $C$ is IL, then
for all $w$ $\in $ $\Sigma ^*$,
$$d_G(w)\geq |\Sigma | ^{|w| -
    |C(w)|-\frac{|w|}{k}(l+\log m+\log k+1)-(km+l+\log m+\log k+1)},$$
where $l=\lceil \log |Q| \rceil$ and $m=$ $\max $ $\{|\nu
(q,b,a)|\mid q\in Q, b\in \Sigma , a\in \Gamma ^2\}$.
\end{lemma}

{\bfseries Proof of Lemma \ref{lemma4}.} Assume the hypothesis, let
$l$ and $m$ be as given, and let $w$ $\in $ $\Sigma ^*$. Fix $0\leq
j<k$ such that $|w|+j$ is divisible by $k$. By Lemma \ref{lemma3} we
have
\begin{equation*}
\begin{aligned}
    d_G(w)&\geq |\Sigma | ^{-j}d_G(w0^j)\\
    &\geq |\Sigma | ^{-j+|w0^j|-|C(w0^j)|-\frac{|w0^j|}{k}(l+\log m+\log k+1)}\\
    &=|\Sigma | ^{|w|-|C(w0^j)|-\frac{|w|}{k}(l+\log m+\log k+1)-\frac{j}{k}(l+\log m+\log k+1)}\\
    &\geq |\Sigma | ^{|w|-|C(w)|-\frac{|w|}{k}(l+\log m+\log k+1)-(km+l+\log m+\log k+1)}
\end{aligned}
\end{equation*}
\begin{flushright}
    $\Box $
    \end{flushright}
Let $l=\lceil \log |Q|\rceil$ and $m=$ $\max \{|\nu (q,b,a)|\mid
q\in Q, b\in \Sigma , a\in \Gamma ^2\}$, and fix $k$ $\in $
$\mathbb{Z}^+$ such that $\frac{l+\log m+\log k+1}{k}<s-s'$. Let
$G=G(C,k)$ be as in Construction \ref{constructioncg}. Then, by
Lemma \ref{lemma4}, for all $n$ $\in $ $I$ we have
$$d^{(s)}_G(w_n)\geq |\Sigma |^{sn-|C(w_n)|-\frac{n}{k}(l+\log m+\log k+1)-(km+l+\log m+\log k+1)}$$
$$\geq |\Sigma |^{(s-s'-\frac{l+\log m+\log k+1}{k})n-(km+l+\log m+\log
    k+1)}$$
Since $s-s'-\frac{l+\log m+\log k+1}{k} >0$, this implies that $S\in
S^\infty [d^{(s)}_G]$.

 Thus, $\dimbpd(S)\leq s$.

 To see that $\robpd(S)\leq \dimbpd(S)$, let
$s>s'>s''>\dimbpd(S)$. It suffices to show that $\robpd(S)\leq s$.
By our choice of $s''$, there is a 1-BPDG $G$ such that the set
$$J=\{n\in \mathbb{N}\mid d^{s''}_G(w_n)\geq 1\}$$
is infinite. By Lemma \ref{gamblers-lemma} there is a nonvanishing
1-BPDG $\widetilde{G}$ such that \newline $d_{\widetilde{G}}(w)\geq
|\Sigma |^{(s''-s')|w|}d_G(w)$ for all $w$ $\in $ $\Sigma ^*$.

\begin{construction} \label{constructiongc} Let $G=(Q, \Sigma , \Gamma
, \delta , \beta , q_0, z_0)$ be a nonvanishing 1-BPDG, and let
$k\in \mathbb{Z}^+$. For each $z\in \Gamma ^*$ (long enough for
$d_{G_{q,z}}(w)$ to be defined for all $w$ $\in$ $\Sigma ^k$) and
$q\in Q$, let $G_{q,z}=(Q, \Sigma , \Gamma , \delta , \beta , q,
z)$, and define $p_{q,z}:\Sigma ^k\rightarrow [0,1]$ by
$p_{q,z}(w)=|\Sigma | ^{-k}d_{G_{q,z}}(w)$. Since G is nonvanishing
and each $d_{G_{q,z}}$ is a martingale with $d_{G_{q,z}}(\lambda
)=1$, each of the functions $p_{q,z}$ is a positive probability
measure on $\Sigma ^k$. For each $z\in \Gamma ^*$, $q\in Q$, let
$\Theta _{q,z}:\Sigma ^k\rightarrow \Sigma ^*$ be the
Shannon-Fano-Elias code given by the probability measure $p_{q,z}$.
Then \\
\\$|\Theta_{q,z}(w)|=l_{q,z}(w)$\\
\\$l_{q,z}(w)=1+\lceil \log \frac{1}{p_{q,z}(w)}\rceil\\
\\$for all  $q \in
Q$ and $w\in \Sigma ^k$, and each of the sets $range(\Theta _{q,z})$
is an instantaneous code. We define the $1$-BPDC $C=C(G,k)=(Q',
\Sigma , \Gamma ', \delta
', \nu ', q_0', z_0')$ whose components are as follows:\\
\\i) $Q'=Q\times \Sigma ^{<k}$\\
\\ii) $q_0'=(q_0,\lambda )$\\
\\iii) $\Gamma '=\bigcup\limits_{i=2k}^{4k-1} \Gamma ^{i}$\\
\\iv) $z_0'=z_0^{2k}$\\
\\v) $\forall (q,w)\in Q'$, $b\in \Sigma $, $a\in \Gamma
'$,\\
\\$\delta '((q,w),b,a)=$
$\left\{\begin{array}{ll}
                          (q,wb,a)  & \hbox{if $|w|<k-1$,} \\
                          (\delta _Q(q,wb,\overline{a}), \lambda , \widehat{\delta _{\Gamma ^*}(q,wb,\overline{a})})
                          & \hbox{if $|w|=k-1$.}
                          \end{array}
                        \right.$\\
\\vi) $\forall (q,w)\in Q'$, $b\in \Sigma $, $a\in
\Gamma ' $,\\
\\$\nu '((q,w),b,a)=$
$\left\{\begin{array}{ll}
                         \lambda  & \hbox{if $|w|<k-1$,} \\
                         \Theta_{q,\overline{a}}(wb) & \hbox{if $|w|=k-1$.}
                       \end{array}
                     \right.$
\end{construction}

Since each range($\Theta _{q,z}$) is an instantaneous code, it is
easy to see that the BPDC $C=C(G,k )$ is IL.

 \begin{lemma} \label{lemma5} In Construction \ref{constructiongc}, if $|w|$ is a
multiple of $k$, then
$$|C(w)|\leq \Big(1+\frac{2}{k}\Big)|w|-\log d_G(w).$$
\end{lemma}

{\bfseries Proof of Lemma \ref{lemma5}.} Let $w=w_0w_1\cdots
w_{n-1}$, where each $w_i$ $\in $ $\Sigma ^k$. For each $0\leq i<n$,
let $q_i=\delta _Q(w_0\cdots w_{i-1})$ and $z_i=\delta _{\Gamma
^*}(w_0\cdots w_{i-1})$. Then,
\begin{equation*}
\begin{aligned}
    |C(w)|&=\sum\limits_{i=0}^{n-1} l_{q_i,z_i}(w_i)\\
    &=\sum\limits_{i=0}^{n-1} \Big(1+ \lceil \log \frac{1}{p_{q_i,z_i}(w_i)}\rceil \Big)\leq \sum\limits_{i=0}^{n-1}
    \Big(2+\log \frac{1}{p_{q_i,z_i}(w_i)}\Big)\\
   & =\sum\limits_{i=0}^{n-1} \bigg(2+\log \frac{|\Sigma |^k}{d_{G_{q_i,z_i}}(w_i)}\bigg)=(k+2)n-\log \prod\limits_{i=0}^{n-1}
   d_{G_{q_i,z_i}}(w_i)\\
    &=(k+2)n-\log d_G(w)=(1+\frac{2}{k})|w|-\log d_G(w)
\end{aligned}
\end{equation*}
\begin{flushright}
    $\Box $
\end{flushright}

\begin{lemma} \label{lemma6} In Construction \ref{constructiongc}, for all $w$ $\in $ $\Sigma ^*$,
$$|C(w)|\leq \Big(1+\frac{2}{k}\Big)|w|-\log d_G(w).$$
\end{lemma}

{\bfseries Proof of Lemma \ref{lemma6}.}
If $|w|$ is multiple of $k$, then we apply the Lemma \ref{lemma5}.\\
Otherwise, let $w=w'z$, where $|w'|$ is a multiple of $k$ and
$|z|=j$, $0<j<k$.\\
\\Then, Lemma \ref{lemma5} tell us that
\begin{equation*}
\begin{aligned}
    |C(w)|&=|C(w')|\\
&\leq \Big(1+\frac{2}{k}\Big)|w'|-\log d_G(w')\\
&\leq \Big(1+\frac{2}{k}\Big)|w'|-\log (|\Sigma |^{-j}d_G(w))\\
&=\Big(1+\frac{2}{k}\Big)|w|-\log d_G(w)-\frac{2j}{k}\\
&\leq \Big(1+\frac{2}{k}\Big)|w|-\log d_G(w).
\end{aligned}
\end{equation*}
\begin{flushright}
    $\Box $
\end{flushright}
Fix $k>\frac{2}{s-s'}$, and let $C=C(\widetilde{G},k)$ be as in
Construction \ref{constructiongc}. Then Lemma \ref{lemma6} tell us
that for all $n$ $\in $ $J$,
\begin{equation*}
\begin{aligned}
    \mid C(w_n)\mid &\leq
    \Big(1+\frac{2}{k}\Big)n-\log d_{\widetilde{G}}(w_n)\\
&\leq \Big(1+\frac{2}{k}+s'-s''\Big)n-\log d_G(w_n)\\
&\leq \Big(\frac{2}{k}+s'\Big)n -\log d^{s''}_G(w_n)\\
&\leq \Big(\frac{2}{k}+s'\Big)n\\
&< sn.
\end{aligned}
\end{equation*}
Thus, $\robpd(S)\leq s$.
\begin{flushright}
    $\Box $
\end{flushright}

\section{Proof of Theorem \ref{LZ}}

For a string $x$, $x^{-1}$ denotes $x$ written in reverse order.

{\bfseries Proof of Theorem \ref{LZ}}  Let $m\in\N$, and let
$k=k(m)$ be an integer to be determined later.
        For any integer $n$, let $T_n$ denote the set of  strings $x$ of size $n$ such that
        $1^j$ does not appear in $x$, for every $j\geq k$.
        Since $T_n$ contains $\bool{k-1}\times \{ 0 \} \times \bool{k-1}\times \{ 0 \} \ldots $
        (i.e. the set of strings whose every $k$th bit is zero),
        it follows that
        $|T_n|\geq 2^{an}$, where $a=1-1/k$.

        \begin{remark} \label{r.extension}
            For every string $x\in T_n$ there is a string $y\in T_{n-1}$ and a bit $b$ such that $yb=x$.
        \end{remark}

        Let $A_n = \{a_1,\ldots a_u\}$ be the set of palindromes in $T_n$. Since fixing the $n/2$ first bits of a palindrome (wlog $n$ is even)
        completely determines it, it follows that $|A_n| \leq 2^{\frac{n}{2}}$.
        Let us separate the remaining strings in $T_n-A_n$ into two sets
        $X_n = \{x_1,\ldots x_t\}$ and $Y_n = \{y_1,\ldots y_t\}$ with
        $(x_i)^{-1}=y_i$ for every $1\leq i \leq t$. Let us choose $X,Y$ such that $x_1$ and $y_t$ start with a zero.
        We construct $S$ in stages. For $n \leq k-1$,
        $S_n$ is an enumeration of all strings of size $n$ in lexicographical order.
        For $n\geq k$,
        $$S_n = a_1 \ldots a_u \  1^{2n} \ x_1 \ldots x_t \ 1^{2n+1} \ y_t \ldots y_1 $$
        i.e. a concatenation of all strings in $A_n$ (the $A$ zone of $S_n$) followed by a flag of $2n$ ones,
        followed by the concatenations of all strings in $X$ (the $X$-zone) and $Y$ (the $Y$ zone)
        separated by a flag of $2n+1$ ones.
        Let $$S=S_1 S_2 \ldots S_{k-1} \ 1^k \ 1^{k+1} \ \ldots 1^{2k-1} \ S_{k} S_{k+1} \ldots $$
        i.e. the concatenation of the $S_j$'s with some extra flags between $S_{k-1}$ and $S_k$.
        We claim that the parsing of $S_n$ ($n\geq k$) by LZ, is as follows:
        $$S_n = a_1, \ldots, a_u, \  1^{2n}, \ x_1, \ldots, x_t, \ 1^{2n+1}, \ y_t, \ldots, y_1 .$$
        Indeed after $S_1, \ldots S_{k-1} \ 1^k \ 1^{k+1} \ \ldots 1^{2k-1}$, LZ has parsed every
        string of size $\leq k-1$ and the flags $1^k \ 1^{k+1} \ \ldots 1^{2k-1}$. Together with Remark \ref{r.extension},
        this guarantees that LZ parses $S_n$ into phrases that are exactly all the strings in $T_n$ and
        the two flags $1^{2n},1^{2n+1}$.

        Let us compute the compression ratio $\rho_{LZ} (S)$.
        Let $n,i$ be integers. By construction of $S$, LZ encodes every phrase in $S_i$ (except the two flags),
          by a phrase in $S_{i-1}$ (plus a bit).
        Indexing a phrase in $S_{i-1}$ requires a codeword of length at least logarithmic in the number of phrase parsed
        before, i.e. $\log (C(S_1 S_2 \ldots S_{i-2}))$.
        Since $C(S_i)\geq |T_i| \geq 2^{ai}$, it follows
        $$
        C(S_1 \ldots S_{i-2}) \geq \sum^{i-2}_{j=1}2^{aj} = \frac{2^{a(i-1)} -2^a}{2^a-1} \geq b 2^{a(i-1)}
        $$
        where $b=b(a)$ is arbitrarily close to $1$.
        Letting $t_i=|T_i|$, the number of bits output by LZ on $S_i$ is at least
        \begin{align*}
        C(S_i) \log C(S_1\ldots S_{i-2})
        &\geq t_i \log b 2^{a(i-1)}\\
        &\geq c t_i(i-1)
        \end{align*}
        where $c=c(b)$ is arbitrarily close to $1$.
        Therefore
        $$
        |LZ(S_1 \ldots S_n)| \geq  \sum_{j=1}^{n} c t_j(j-1)
        $$
        Since $|S_1 \ldots S_n| \leq 2k^2 + \sum_{j=1}^{n} (j t_j + 4j)  $, (the two flags plus the extra flags between
            $S_{k-1}$ and $S_k$) the compression ratio is given by
        \begin{align}
        \rho_{LZ}(S_1 \ldots S_n)   &\geq c\frac{\sum_{j=1}^{n} t_j(j-1) }{2k^2+\sum_{j=1}^{n} j (t_j+4)}\\
        &= c - c \frac{2k^2+ \sum_{j=1}^{n} (t_j+4j)}{2k^2+\sum_{j=1}^{n} j (t_j+4)} \label{e.secondterm}
        \end{align}
        The second term in Equation \ref{e.secondterm} can be made arbitrarily small for $n$ large enough:
        Let $M\leq n$, we have
        \begin{align*}
        2k^2+\sum_{j=1}^{n} j (t_j+4) &\geq 2k^2+\sum_{j=1}^{M} j t_j + (M+1)\sum_{j=M+1}^{n}  t_j\\
        &= 2k^2+ \sum_{j=1}^{M} j t_j + M\sum_{j=M+1}^{n} t_j+ \sum_{j=M+1}^{n}  t_j\\
          &\geq 2k^2+ \sum_{j=1}^{M} j t_j + M\sum_{j=M+1}^{n} t_j+ \sum_{j=M+1}^{n}  2^{aj}\\
          &\geq 2k^2+ \sum_{j=1}^{M} j t_j + M\sum_{j=M+1}^{n} t_j+ 2^{an}\\
          &\geq M \sum_{j=M+1}^{n} t_j+ M(2k^2 + 2n(n+1) + \sum_{j=1}^{M}t_j)\quad \text{for $n$ big enough}\\
          &= M (2k^2+\sum_{j=1}^{n} t_j+4\sum_{j=1}^{n}j)
        \end{align*}
        Hence
        $$
        \rho_{LZ}(S_1 \ldots S_n)   \geq c - \frac{c}{M}
        $$
        which by definition of $c,M$ can be made arbitrarily close to $1$ by choosing $k$ accordingly, i.e
        $$
        \rho_{LZ}(S_1 \ldots S_n)   \geq 1- \frac{1}{m}.
        $$

        Let us show that $\dimbpd(S)\leq \frac{1}{2}$. Consider the following BPD martingale $d$. Informally,
        $d$ on $S_n$ goes through the $A_n$ zone until the first flag, then starts pushing the whole $X$
        zone onto its stack until it hits the second flag. It then uses the stack to bet correctly on the
        whole $Y$ zone. Since the $Y$ zone is exactly the $X$ zone written in reverse order, $d$ is able to
        double its capital on every bit of the $Y$ zone. On the other zones, $d$ does not bet. Before giving a detailed
        construction of $d$, let us compute the upper bound it yields on $\dimbpd(S)$.
        \begin{align*}
        \dimbpd(S)
        &\leq 1 - \limsup_{n\rightarrow\infty} \frac{\log d(S_1\ldots S_n)}{|S_1\ldots S_n|}\\
        &\leq 1 - \limsup_{n\rightarrow\infty} \frac{ \sum_{j=1}^{n} |Y_j| }{ 2k^2+\sum_{j=1}^{n} (j|T_j|+4j) }\\
        &\leq 1 - \limsup_{n\rightarrow\infty} \frac{ \sum_{j=1}^{n} j\frac{|T_j|-|A_j|}{2} }{  2k^2+\sum_{j=1}^{n} (j|T_j|+4j) }\\
        &\leq \frac{1}{2} + \frac{1}{2} \limsup_{n\rightarrow\infty}  \frac{2k^2 +\sum_{j=1}^{n} (j|A_j|+4j) }{ 2k^2+\sum_{j=1}^{n} (j|T_j|+4j)}.
        \end{align*}
        Since
        \begin{align*}
        \limsup_{n\rightarrow\infty}  \frac{2k^2 +\sum_{j=1}^{n} (j|A_j|+4j) }{ 2k^2+\sum_{j=1}^{n} (j|T_j|+4j)}
        &\leq       \limsup_{n\rightarrow\infty}  \frac{\sum_{j=1}^{n} j(|A_j|+4 +2k^2 )}{ \sum_{j=1}^{n} |T_j|}\\
        &\leq       \limsup_{n\rightarrow\infty}  \frac{\sum_{j=1}^{n} j(2^{\frac{j}{2}} + 2^{\frac{j}{4}} )}{ \sum_{j=1}^{n} 2^{aj}}\\
          &\leq \limsup_{n\rightarrow\infty}\frac{n2^{\frac{3n}{4}}}{2^{an}}\\
        &= 0 .
        \end{align*}
        It follows that
        $$\dimbpd(S) \leq \frac{1}{2}.$$

        Let us give a detailed description of $d$. Let $Q$ be the following set of states:
        \begin{itemize}
            \item   The start state $q_0$, and $q_1,\ldots q_v$ the ``early'' states that will count up to
                    $$v=|S_1 S_2 \ldots S_{k-1} \ 1^k \ 1^{k+1} \ \ldots 1^{2k-1}|.$$
            \item   $q^a_0, \ldots, q^a_k$ the $A$ zone states that cruise through the $A$ zone until the first flag.
            \item   $q^{1f}$ the first flag state.
            \item   $q^X_0, \ldots, q^X_k$ the $X$ zone states that cruise through the $X$ zone, pushing every bit on the stack,
                    until the second flag is met.
            \item   $q^r_0, \ldots, q^r_k$ which after the second flag is detected, pop $k$ symbols from the stack that were
                    erroneously pushed while reading the second flag.
            \item $q^{2f}$ the second flag state.
            \item $q^b$ the betting on zone $Y$ state.
        \end{itemize}
        Let us describe the transition function $\delta :Q \times \bool{} \times\bool{}\rightarrow Q\times\bool{}$.
        First $\delta$ counts until $v$ i.e. for $i=0,\ldots v-1$
        $$
            \delta(q_i,x,y) = (q_{i+1},y) \quad \text{ for any } x,y
        $$
        and after reading $v$ bits, it enters in the first $A$ zone state, i.e. for any $x,y$
        $$\delta(q_v,x,y) = (q^a_0,y).$$
        Then $\delta$ skips through $A$ until the string $1^k$ is met, i.e. for $i=0,\ldots k-1$ and any $x,y$
        $$
            \delta(q^a_i,x,y) =
        \begin{cases}
        (q^a_{i+1},y) &\text{ if } x=1\\
        (q^a_{0},y) &\text{ if } x=0\\
        \end{cases}
        $$ and
        $$
        \delta(q^a_k,x,y) = (q^{1f},y).
        $$
        Once $1^k$ has been seen, $\delta$ knows the first flag has started, so it skips
        through the flag until a zero is met, i.e. for every $x,y$
        $$
            \delta(q^{1f},x,y) =
        \begin{cases}
        (q^{1f},y) &\text{ if } x=1\\
        (q^X_{0},0y) &\text{ if } x=0\\
        \end{cases}
        $$
        where state $q^X_0$ means that the first bit of the $X$ zone (a zero bit) has been read, therefore $\delta$ pushes a zero.
        In the $X$ zone, delta pushes every bit it sees until it reads a sequence of $k$ ones, i.e until the start of the second flag, i.e
        for $i=0,\ldots k-1$ and any $x,y$
        $$
            \delta(q^X_i,x,y) =
        \begin{cases}
        (q^X_{i+1},xy) &\text{ if } x=1\\
        (q^X_{0},xy) &\text{ if } x=0\\
        \end{cases}
        $$
        and
        $$
        \delta(q^X_k,x,y) = (q^{r}_0,y).
        $$
        At this point, $\delta$ has pushed all the $X$ zone on the stack, followed by
        $k$ ones. The next step is to pop $k$ ones,
        i.e
        for $i=0,\ldots k-1$ and any $x,y$
        $$
            \delta(q^r_i,x,y) = (q^r_{i+1},\lambda)
        $$
        and
        $$
        \delta(q^r_k,x,y) = (q^{2f}_0,y).
        $$
        At this stage, $\delta$ is still in the second flag (the second flag is always bigger than $2k$)
        therefore it keeps on reading ones until a zero (the first bit of the $Y$ zone) is met. For any $x,y$
        $$
            \delta(q^{2f},x,y) =
        \begin{cases}
        (q^{2f},y) &\text{ if } x=1\\
        (q^b,\lambda) &\text{ if } x=0 .
        \end{cases}
        $$
        On the last step $\delta$ has read the first bit of the $Y$ zone, therefore it pops it. At this stage,
        the stack  exactly contains the $Y$ zone (i.e. the $X$ zone written in reverse order) except the first bit;
        $\delta$ thus uses its stack to bet and double its capital on every bit in the $Y$ zone. Once the stack is empty,
        a new $A$ zone begins. Thus, for any $x,y$
        $$
        \delta(q^b,x,y) = (q^b,\lambda).
        $$
        and
        $$
            \delta(q^{b},x,z_0) =
        \begin{cases}
        (q^a_1,z_0) &\text{ if } x=1\\
        (q^a_0,z_0) &\text{ if } x=0 .
        \end{cases}
        $$
        The betting function is equal to $1/2$ everywhere (i.e no bet) except on state $q^b$,
        where
        $$
            \beta(q^{b},y)(z) =
        \begin{cases}
        1 &\text{ if } y=z\\
        0 &\text{ if } y\neq z .
        \end{cases}
        $$
        and $\beta$ stops betting once start stack symbol is met, i.e.
        $$
        \beta(q^b,z_0) = \frac{1}{2}.
        $$
        \qed


\end{document}